\documentclass[11pt,a4paper,reqno,twoside]{amsart}
\usepackage{amssymb}
\usepackage{amsthm}
\usepackage{xspace}
\usepackage{amsmath}
\usepackage{setspace}
\usepackage{amscd}
\usepackage{natbib}
\usepackage{setspace}
\usepackage{enumerate}
\usepackage[english]{babel}
\usepackage{algorithmic,algorithm}

\newtheorem{theorem}{Theorem}[section]
\newtheorem{lemma}[theorem]{Lemma}
\newtheorem{proposition}[theorem]{Proposition}
\newtheorem{corollary}[theorem]{Corollary}
\newtheorem{definition}[theorem]{Definition}

\newtheorem{example}[theorem]{Example} 
       
\begin{document}
\title[Polynomial Rings over Rings]
{Reduced Gr\"obner Bases and Macaulay-Buchberger Basis Theorem over Noetherian Rings}
\author[Maria Francis and  Ambedkar Dukkipati]{Maria Francis and  Ambedkar Dukkipati}

\email{mariaf@csa.iisc.ernet.in \\ ad@csa.iisc.ernet.in}
\address{Dept. of Computer Science \& Automation\\Indian Institute of Science, Bangalore - 560012}
%
\maketitle
\begin{abstract}
In this paper, we extend the characterization of $\mathbb{Z}[x]/\langle f \rangle$, where $f \in \mathbb{Z}[x]$ to be a free $\mathbb{Z}$-module to multivariate polynomial rings
over any commutative Noetherian ring, $A$. 
The characterization allows us to extend 
the Gr\"obner basis method of computing a $\Bbbk$-vector space basis of residue class polynomial rings over a field $\Bbbk$ (Macaulay-Buchberger Basis Theorem) to 
rings, i.e. $A[x_1,\ldots,x_n]/\mathfrak{a}$, 
where $\mathfrak{a} \subseteq A[x_1,\ldots,x_n]$ is an ideal. 
We give some insights into the characterization for two special cases, when $A = \mathbb{Z}$ and $A = \Bbbk[\theta_1,\ldots,\theta_m]$. 
As an application of this characterization, we show that the concept of border bases can be extended to rings when the corresponding residue class ring is a finitely generated, free $A$-module.

\end{abstract}

\section{Introduction}

\label{Introduction} 

\noindent\cite{Buchberger:1965:thesis} introduced the algorithmic theory of Gr\"obner bases and gave an algorithm for finding a $\Bbbk$-vector space basis of the residue class ring of a zero dimensional ideal. 
Since then the theory of Gr\"obner bases has become a standard tool in computational ideal theory and algebraic
geometry. Subsequently, the theory of Gr\"obner bases has been extended to different variations of the polynomial ring. 
The variants include polynomial rings over rings \citep{Zacharias:1978:grobnerbasisrings3}, monoid rings \citep{Madlener93ongrobner,kreuzer2006grobnerbasiscrypto}, 
free associative algebras \citep{Eisenbud:1998:freeassociativealgebra}, etc.
This paper deals with some aspects of Gr\"obner bases of polynomial rings over commutative Noetherian rings.

To extend Gr\"obner bases theory for polynomial rings over a ring $A$, various approaches have been proposed
\cite[e\,.g.][]{Trinks:1978:grobnerforrings1, Moller:1988:grobnerrings2, Zacharias:1978:grobnerbasisrings3}.  
For a good exposition on Gr\"obner bases over rings one can refer to \citep{Adams:1994:introtogrobnerbasis}. 
However, these approaches only looked at extending basic definitions and concepts over rings, and validity of many important results were not explored \citep{Greuel:2011:idealvanishingpolynomials}.
Recently, there has been renewed interest in polynomial rings over rings \citep{Greuel:2011:idealvanishingpolynomials}.
For instance, certain residue class rings over $\mathbb{Z}[x]$ called ideal lattices \citep{Micciancio:2002:CyclicLattices} have 
 shown to be isomorphic to integer lattices, an important cryptographic primitive \citep{Ajtai:1996:Znascryptoprimitive}
and certain cyclic lattices in $\mathbb{Z}[x]$ have been used in NTRU cryptographic schemes \citep{Hoffstein:1998:NTRUconf}. 
Boolean polynomial rings over a boolean ring have been used to solve Sudoku and other combinatorial puzzles \citep{Sato:2011:BooleanGB}.
Further, polynomial rings over $\mathbb{Z}/2^k$ have been used to prove the correctness of data paths in system-on-chip design \citep{Greuel:2011:idealvanishingpolynomials}.  
Also, the widely used degree truncated polynomial rings are
actually the quotient rings of the form, $\mathbb{Z}[x_1,\ldots,x_n]/\langle x^{r_1}-1,\cdots, x^{r_n}-1\rangle$, where each $r_i$ is a positive integer. 

As mentioned above, ideal lattices are integer lattices that are ideals as well in certain residue class polynomial rings over $\mathbb{Z}$. For all
ideals in a residue class ring to be lattices the ring itself should be isomorphic to an integer lattice. In $\mathbb{Z}[x]$, the necessary and sufficient
condition for the quotient ring $\mathbb{Z}[x]/\langle f \rangle$ to be isomorphic to $\mathbb{Z}^n$, where $f\in \mathbb{Z}[x]$ and $n$ is the degree of the polynomial, 
is that $f$ should be a monic polynomial. Ideal lattices in $\mathbb{Z}[x]$ with an extra condition that $f$ should be an irreducible polynomial, are used in cryptography. 
Finding an approximate shortest vector is hard in these algebraic structures making them a good choice to build efficient collision-resistant hash functions \citep{Lyubashevsky:2008:Ideallattice}. 
The question we ask here is that how to characterize  an ideal $\mathfrak{a} \subseteq \mathbb{Z}[x_1,\ldots,x_n]$ such that all ideals in $\mathbb{Z}[x_1,\ldots,x_n]/\mathfrak{a}$ are isomorphic to 
integer lattices, i.e. $\mathbb{Z}[x_1,\ldots,x_n]/\mathfrak{a}$ is free. In this paper, we study the more 
general problem of characterizing residue class polynomial rings over arbitrary rings as free modules. This characterization involves reduced Gr\"obner basis over rings. 
\subsection*{Contributions}
One of the recent works in the theory of Gr\"obner bases of polynomial rings over rings has been the extension of the concept of reduced Gr\"obner bases to polynomial rings 
over arbitrary rings \citep{Pauer:2007:Grobnerbasisrings}
and over polynomial rings in particular \citep{Nabeshima:2009:polyringoverpolyring}. 
We use the definition of reduced Gr\"obner basis \citep{Pauer:2007:Grobnerbasisrings} and a restriction
on the ideals in the coefficient ring $A$ to define a basis called `short reduced Gr\"obner basis' 
to arrive at a necessary and sufficient condition for a finitely generated  $A[x_1,\ldots,x_n]/\mathfrak{a}$ to have a free representation w.r.t. the Gr\"obner basis. 
We then state the Macaulay-Buchberger basis theorem for a free $A[x_1,\ldots,x_n]/\mathfrak{a}$ which gives a Gr\"obner basis algorithm  to determine an $A$-module basis for $A[x_1,\ldots,x_n]/\mathfrak{a}$. 

We look at the characterization for two special rings, 
$A= \mathbb{Z}$ and $A= \Bbbk[\theta_1,\ldots,\theta_m]$. 
In the case of $A= \Bbbk[\theta_1,\ldots,\theta_m]$, we look at another definition of reduced Gr\"obner basis given by \cite{Nabeshima:2009:polyringoverpolyring} 
called strong reduced Gr\"obner basis and prove the characterization in terms of this definition as well. We also show that the short reduced Gr\"obner basis is the same as the
strong reduced Gr\"obner basis when $A = \Bbbk[\theta_1,\ldots,\theta_m]$.

An important application of the
 characterization is that we can 
directly extend the concept of border bases previously defined for zero-dimensional ideals in $\Bbbk[x_1,\ldots,x_n]$ to ideals in $A[x_1,\ldots,x_n]$ which 
satisfy the conditions given in the characterization.
\subsection*{Organization}
The rest of the paper is organized as follows: 
In Section ~\ref{Preliminaries}, we recall briefly Macaulay-Buchberger theorem over fields. 
In Section ~\ref{Characterization}, we give a necessary and sufficient condition for the quotient ring $A[x_1,\ldots,x_n]/\mathfrak{a}$ to have a free $A$-module representation w.r.t. a Gr\"obner basis. 
In Section ~\ref{M-B}, we give the Macaulay-Buchberger basis theorem for a free $A$-module, $A[x_1,\ldots,x_n]/\mathfrak{a}$, along with an algorithm to compute
an $A$-module basis. We study two special cases of the coefficient ring, $A$ in Section ~\ref{specialcases}. 
In Section ~\ref{borderbases}, we extend Border basis to $A[x_1,\ldots,x_n]$ directly,
the characterization enables us to do so. 
\section{Background \& Preliminaries}
\label{Preliminaries}
Throughout this paper, $\Bbbk$ denotes a field, $A$ a Noetherian commutative ring,  $\mathbb{Z}$ the ring of integers and $\mathbb{N}$ the set of positive integers including zero. 
A polynomial ring in indeterminates $x_1,\ldots,x_n$ over $A$ is denoted as $A[x_1,\ldots,x_n]$.  
We represent a monomial in $x_1,\ldots, x_n$ (or $\theta_1,\ldots,\theta_n$) as $x^{\alpha}$ (or $\theta^{\alpha}$) where  $\alpha\in {\mathbb{Z}}_{\ge 0}^n$.  
The monoid isomorphism between the set of all monomials in indeterminates $x_1,\ldots, x_n$ and  ${\mathbb{Z}}_{\ge 0}^n$ allows us to denote the set of all monomials as ${\mathbb{Z}}_{\ge 0}^n$.  
We assume that there is a monomial order $\prec$ on the monomials in the indeterminates $x_{1},\ldots,x_{n}$. 
With respect to this monomial order, we have the leading monomial ($\mathrm{lm}_\prec$), leading coefficient ($\mathrm{lc}_\prec$), leading term ($\mathrm{lt}_\prec$) and degree of a
polynomial ($\mathrm{deg}_\prec$), 
where $\mathrm{lt}_\prec(f) = \mathrm{lc}_\prec(f)\mathrm{lm}_\prec(f)$ and $\mathrm{deg}_\prec(f) = \mathrm{deg}_\prec(\mathrm{lm}_\prec(f))$ in $A[x_{1},\ldots,x_{n}]$. 
With this notation, the leading term ideal (or initial ideal) of a set $S \subseteq A[x_{1},\ldots,x_{n}]$, is  $\langle\mathrm{lt}_\prec(S) \rangle = \langle \{\mathrm{lt}_\prec(f) \mid f \in S \} \rangle$. 
When there is no confusion regarding which monomial order to consider we omit the monomial order subscript $\prec$ from the notations.\\
\par Here, we recall some definitions and the Macaulay-Buchberger Basis theorem. 
\begin{definition}
Let $\mathfrak{a} \subseteq \Bbbk[x_1,\ldots,x_n]$ be an ideal. We call a monomial $x^\alpha$ in $\Bbbk[x_1,\ldots,x_n]$, 
a standard monomial w.r.t. $\mathfrak{a}$ if none of the leading terms of the ideal divide the monomial, i.e. $x^\alpha \notin \langle \mathrm{lt}(\mathfrak{a})\rangle$. 
\end{definition}
\begin{corollary}
Let $G$ be a Gr\"obner basis for an ideal $\mathfrak{a}\subseteq \Bbbk[x_1,\ldots,x_n]$.  
A monomial, $x^\alpha$ is a standard monomial w.r.t. $\mathfrak{a}$ if and only if  $x^\alpha \notin \langle \mathrm{lt}(G)\rangle$.
\end{corollary}
\begin{theorem}[Macaulay Basis Theorem \citep{kreuzer2000computational}]
 Let $\mathfrak{a}$ be an ideal in $\Bbbk[x_1,\ldots,x_n]$. The residue classes of the terms in $\Bbbk[x_1,\ldots,x_n]/\langle \mathrm{lt}(\mathfrak{a})\rangle$ 
form a $\Bbbk$-vector space basis of $\Bbbk[x_1,\ldots,x_n]/\langle \mathfrak{a} \rangle$.  That is, the $\Bbbk$-vector space basis  of $\Bbbk[x_1,\ldots,x_n]/\langle \mathfrak{a} \rangle$ is $S = \{ x ^\alpha + \mathfrak{a} :  x^\alpha 
\notin \langle \mathrm{lt}(\mathfrak{a})\rangle\}$. 
\end{theorem}
The theory of Gr\"obner bases gives us an algorithmic method to determine the $\Bbbk$-vector space basis of $\Bbbk[x_1,\ldots,x_n]/\langle \mathfrak{a} \rangle$. 
Therefore, the Macaulay Basis theorem can be stated in the following manner as well and we refer to it as the Macaulay-Buchberger Basis Theorem in $\Bbbk[x_1,\ldots,x_n]$.
\begin{theorem}[Macaulay-Buchberger Basis Theorem \citep{Buchberger:1965:thesis}]
Let $G = \{g_1, \ldots, g_t\}$ be a Gr\"obner basis for an ideal $\mathfrak{a} \subseteq \Bbbk[x_1,\ldots,x_n]$. 
A basis for the vector space $\Bbbk[x_1,\ldots,x_n]/\langle \mathfrak{a} \rangle$ is given by $S = \{ x ^\alpha + \mathfrak{a} : \mathrm{lm}(g_i)\nmid x^\alpha, i=1,\ldots,t\}$.
 
\end{theorem}

\section{Characterization of finitely generated $A[x_{1},\ldots,x_{n}]/\mathfrak{a}$ as a free $A$ - module}
\label{Characterization}
\noindent One can extend the definition of Gr\"obner bases in the case of polynomial rings over fields to rings. (Reader can refer to the exposition given in 
\citep[Chapter 4, Sections 4.1, 4.2 and 4.3.1]{Adams:1994:introtogrobnerbasis}). 
One can arrive at a definition of reduced Gr\"obner bases over rings analogous to that of fields, as defined in \citep{Arnold:2003:ModularAlgoGrobnerbasis}, but it may not exist in all cases. 
A new definition of reduced Gr\"obner basis over rings is given by \cite{Pauer:2007:Grobnerbasisrings}, and it ensures the existence of a reduced Gr\"obner basis for any ideal in a polynomial ring over the ring $A$. 
Henceforth, ``reduced Gr\"obner basis" refers to ``Pauer's definition of reduced Gr\"obner basis" unless otherwise stated. Before we proceed further we give a brief account of this concept. 
\subsection{Reduced Gr\"obner Bases over Rings \citep{Pauer:2007:Grobnerbasisrings}}
We introduce the following notations and definitions. 
For any ideal $I$ in $A$, we select a finite system, $\mathrm{Gen}(I)$ of generators of $I$, and 
a mapping  $\eta _I$ from $A$ to $A$ such that $\eta _I(0)= 0$, $\eta _I$ is constant for each coset of $I$ 
and for any $z \in A$ we have $\eta _I(z) \in z+ I$.
\begin{example}
 Let $A=\mathbb{Z}$. Let $I$ be an ideal generated by $a_1,\ldots,a_m$ and $a = \mathrm{gcd}(a_1,\ldots,a_m)$. Let $z \in \mathbb{Z}$. 
 Then we can choose $\mathrm{Gen} (I) = \{a\}$ and $\eta _I(z) = z\hspace{2pt} \mathrm{mod} \hspace{2pt}a$.
\end{example}
Let $\mathfrak{a}$ be an ideal in $A[x_{1},\ldots,x_{n}]$ and $\alpha \in \mathbb{N}^n$.  Let $G$ be a Gr\"obner basis for $\mathfrak{a}$ and let 
$\mathrm{lm}(G)$ denote the set of leading monomials in $G$.
We represent the leading coefficient ideal of all polynomials in  $\mathfrak{a}$ of degree $\alpha$ as $\langle\mathrm{lc}(\alpha,\mathfrak{a})\rangle $, 
i.e. $\langle\mathrm{lc}(\alpha,\mathfrak{a})\rangle = \langle \mathrm{lc}(f):f \in \mathfrak{a}, \mathrm{deg}(f) = \alpha \rangle$. 
Similarly, the leading coefficient ideal of all polynomials in  $\mathfrak{a}$ such that the leading monomial of the polynomials divide $x^\alpha$ is denoted as $\langle\mathrm{lc}(<\alpha,\mathfrak{a})\rangle$. 
We have $\langle\mathrm{lc}(<\alpha,\mathfrak{a})\rangle := \langle \mathrm{lc}(f) : f\in \mathfrak{a}, \alpha \in \mathrm{deg}(f) + \mathbb{N}^n, \alpha \ne \mathrm{deg}(f) \rangle$. 
We use $\mathrm{Gen} (\alpha,\mathfrak{a}) $ to represent the set of all non-zero $\eta_{\langle\mathrm{lc}(<\alpha,\mathfrak{a})\rangle}(a)$,\
 where $a$ belongs to the set of all generators of 
$\langle\mathrm{lc}(\alpha,\mathfrak{a})\rangle $.  
We give below the formal definition of $\mathrm{Gen} (\alpha, \mathfrak{a})$.
\begin{definition}
Let $\mathfrak{a}$ be an ideal in $A[x_{1},\ldots,x_{n}]$ and $\alpha \in \mathbb{N}^n$. Let $G$ be a Gr\"obner basis for $\mathfrak{a}$ and let 
$\mathrm{lm}(G)$ denote the set of leading monomials in $G$. Let $\langle\mathrm{lc}(\alpha, \mathfrak{a})\rangle$ be the leading coefficient ideal of all polynomials in  $\mathfrak{a}$ of degree $\alpha$ and $\langle\mathrm{lc}(<\alpha,\mathfrak{a})\rangle$ be the leading coefficient ideal of all polynomials in  $\mathfrak{a}$ such that the leading monomial of the polynomials divide $x^\alpha$. For each $x^\alpha\in \mathrm{lm}(G)$ we 
define,   
 \begin{displaymath}
 \mathrm{Gen} (\alpha, \mathfrak{a}) = \{\eta_{\langle\mathrm{lc}(<\alpha,\mathfrak{a})\rangle}(a) : a \in \mathrm{Gen}(\langle\mathrm{lc}(\alpha,  \mathfrak{a})\rangle)\} \setminus\{0\}.
 \end{displaymath}
As defined above, $\eta_{\langle\mathrm{lc}(<\alpha,\mathfrak{a})\rangle}(a) $ is an element in the coset, $a + \langle\mathrm{lc}(<\alpha,\mathfrak{a})\rangle$.
\end{definition}
We proceed now to Pauer's definition of reduced Gr\"obner basis over rings.
\begin{definition}\citep{Pauer:2007:Grobnerbasisrings}
 A  Gr\"obner basis $G$ of $\mathfrak{a} \subseteq A[x_{1},\ldots,x_{n}]$ is a reduced Gr\"obner basis w.r.t. a monomial order $\prec$ iff 
\begin{enumerate}[(i)]
 \item for all $\alpha \in {\mathbb{Z}}_{\ge 0}^n$ such that $x^\alpha \in \mathrm{lm}(G)$, the map
 \begin{align*}
   \{g\in G : \mathrm{deg}(g) = \alpha \} &\longrightarrow \mathrm{Gen} (\alpha,\mathfrak{a})\\
                                                             g&\longmapsto \mathrm{lc}(g)
\end{align*}
is bijective and
\item for all $ g:=\sum\limits_{\beta \in \mathbb{N}^n} c_{\beta,g}x^\beta \in G$ and all $\alpha \in \mathbb{N}^n$ with $\alpha \ne \mathrm{deg}(g)$ and $c_{\alpha,g} \ne 0$ we have $c_{\alpha,g} = \eta_{\langle\mathrm{lc}(\alpha,\mathfrak{a})\rangle}(c_{\alpha,g})$.
\end{enumerate}
\end{definition}
\begin{theorem} \citep{Pauer:2007:Grobnerbasisrings}
 There exists a reduced Gr\"obner basis for every ideal $\mathfrak{a} \subseteq A[x_{1},\ldots,x_{n}]$.  
\end{theorem}
It can be seen that different choices of the generators for the leading coefficient ideal of each leading monomial in $G$, $\mathrm{Gen}(\langle\mathrm{lc}(\alpha,  \mathfrak{a} )\rangle)$, lead to different $\mathrm{Gen} (\alpha, \mathfrak{a})$,  which in turn lead to different
reduced Gr\"obner bases.  Once we fix $\mathrm{Gen} (\alpha, \mathfrak{a})$ for all $x^\alpha \in \mathrm{lm}(G)$, the reduced Gr\"obner basis $G$ is unique. 
\begin{theorem}\citep{Pauer:2007:Grobnerbasisrings}
The reduced Gr\"obner basis $G$ for an ideal $\mathfrak{a}  \subseteq A[x_{1},\ldots,x_{n}]$ is unique upto $\mathrm{Gen} (\alpha, \mathfrak{a})$ for all $x^\alpha \in \mathrm{lm}(G)$.
\end{theorem}
Later in this paper, we show that the minimality of the length of the generating set, $\mathrm{Gen} (\alpha, \mathfrak{a})$ for each leading monomial $x^\alpha$ in $G$ is necessary for characterizing a finitely generated $A[x_1,\ldots,x_n]/\mathfrak{a}$ as a free $A$-module. 
 We call such a reduced Gr\"obner basis as `short reduced Gr\"obner basis'. 
\begin{example}\label{Example}
Consider the ideal $\mathfrak{a}$ for which $G= \{ 3 x^2, 5x^2,y\}$ is a Gr\"obner basis. Let us calculate a  short reduced Gr\"obner basis. 
For the leading monomial $x^2$, $\mathrm{Gen}((2,0),\mathfrak{a}) = \{\mathrm{gcd}(3,5) \}= \{1\}$ is the generating set of minimal length.
For the leading monomial $y$,  $\mathrm{Gen}((0,1),\mathfrak{a}) =\{1\}$ is the  generating set of minimal length. 
The short reduced Gr\"obner basis for the ideal is therefore $G= \{x^2,y\}$. \\
Now for the same ideal $\mathfrak{a}$,  let us assume that $\mathrm{Gen}(\langle\mathrm{lc}(\alpha, \mathfrak{a} )\rangle)$ is taken as the same set of generators given in the basis and not their $\mathrm{gcd}$. 
Therefore, for the leading monomial $x^2$, $\mathrm{Gen}((2,0),\mathfrak{a}) = \{3,5\}$ and for the leading monomial $y$,  $\mathrm{Gen}((0,1),\mathfrak{a}) =\{1\}$. 
For each degree $\alpha \in \{(2,0),(0,1)\}$, if we look at the map between $\{g\in G : \mathrm{deg}(g) = \alpha\}$ and $\mathrm{Gen} (\alpha,\mathfrak{a})$ given by each element $g$ mapping to its leading coefficient, it is a bijective map. 
Therefore $G= \{ 3 x^2, 5x^2,y\}$ is a reduced Gr\"obner basis w.r.t. this definition of $\mathrm{Gen}(\langle\mathrm{lc}(\alpha, \mathfrak{a} )\rangle)$. 
Thus, $\mathrm{Gen} (\alpha, \mathfrak{a})$ is a factor that determines the reduced Gr\"obner basis. 
\end{example}

\subsection{Characterization}\label{homomorphism}
Consider an ideal $\mathfrak{a} \subseteq A[x_1,\ldots,x_n]$.  Let $G = \{g_i: i = 1, \ldots, t\}$ be a Gr\"obner basis for $\mathfrak{a}$ w.r.t. a monomial order, $\prec$. 
Recall that, $J_{x^{\alpha}} = \{i : \mathrm{lm}(g_i)\mid x^{\alpha},  g_i \in G \}$  
and $I_{J_{x^{\alpha}}} = \langle \{\mathrm{lc}(g_i) : i \in J_{x^{\alpha}}\} \rangle$ \cite[Chapter 4, Section 4.2 and 4.3.1]{Adams:1994:introtogrobnerbasis}. 
We refer to $I_{J_{x^{\alpha}}}$ as the leading coefficient ideal
w.r.t. $G$.
Consider $A/I_{J_{x^{\alpha}}}$.  We assume that the coefficient ring $A$ has effective coset representatives \cite[Chapter 4, Page 226]{Adams:1994:introtogrobnerbasis}. 
Let $C_{J_{x^{\alpha}}}$ represent a set of coset representatives of the equivalence classes in  $A/I_{J_{x^{\alpha}}}$.
Let $f \in  A[x_1,\ldots,x_n]$. On reducing $f$ with $G$ we get $f = \sum\limits_{i=1}^m a_i x^{\alpha_i} \text{ mod } \langle G \rangle$, where $a_i \in A$. 
If $A[x_{1},\ldots,x_{n}]/\langle G \rangle$ is a finitely generated $A$-module of size $m$, 
then corresponding to coset representatives, $C_{J_{x^{\alpha_1}}}, \ldots, C_{J_{x^{\alpha_m}}}$ and $G$, there exists an $A$-module homomorphism,
\begin{equation} \label{equation}
\begin{split}
 \phi :  A[x_{1},\ldots,x_{n}]/\langle G \rangle &\longrightarrow A/I_{J_{x^{\alpha_1}}} \times \cdots \times A/I_{J_{x^{\alpha_m}}}\\
         \sum\limits_{i=1}^m a_i x^{\alpha_i} + \langle G \rangle &\longmapsto (c_1 +I_{J_{x^{\alpha_1}}}  , \cdots, c_m + I_{J_{x^{\alpha_m}}}) ,
\end{split}
\end{equation}
where $c_i = a_i \text{  mod  } I_{J_{x^{\alpha_i}}}$ and  $c_i \in C_{J_{x^{\alpha_i}}}$. Note that $\phi$ depends on the choice of coset representatives, 
$C_{J_{x^{\alpha_1}}}, \ldots, C_{J_{x^{\alpha_m}}}$ and the monomial order, $\prec$. 

Given a Gr\"obner basis $G$ for 
$\mathfrak{a}$ and the set of coset representatives $C_J$ for the saturated subsets $J$, every $f \in  A[x_1,\ldots,x_n]$ has a unique normal form 
\cite[Chapter 4, Theorem 4.3.3.]{Adams:1994:introtogrobnerbasis}.
The mapping $\phi$ is  surjective by construction.  Consider $f + \mathfrak{a}, g + \mathfrak{a}$ where $f,g \in  A[x_1,\ldots,x_n]$.
On reducing $f,g$ with $G$, we get $f = \sum\limits_{i=1}^m a_i x^{\alpha_i} \text{ mod } \langle G \rangle$ and 
$g = \sum\limits_{i=1}^m b_i x^{\alpha_i} \text{ mod } \langle G \rangle$, where $a_i,b_i  \in A$. 
Let $\phi(f) = (c_1 +I_{J_{x^{\alpha_1}}}  , \cdots, c_m + I_{J_{x^{\alpha_m}}}) $ 
and $\phi(g)=(d_1+I_{J_{x^{\alpha_1}}}  , \cdots, d_m+ I_{J_{x^{\alpha_m}}} )$, where $c_i = a_i \text{  mod  } I_{J_{x^{\alpha_i}}}$,  $d_i=b_i \text{  mod  } I_{J_{x^{\alpha_i}}}$ 
and  $c_i,d_i\in C_{J_{x^{\alpha_i}}}$. Let $c_i = d_i \text{  mod  } I_{J_{x^{\alpha_i}}}$ for all $i \in \{1,\ldots,m\}$. Then $c_i = d_i$ since the set 
of coset representatives,  $C_{J_{x^{\alpha_i}}}$ is fixed.  This implies,  $f - g$ is an 
element of $\mathfrak{a}$ and $f + \mathfrak{a} = g + \mathfrak{a}$. Hence, $\phi$ is injective and an $A$-module isomorphism.

We refer to $A/I_{J_{x^{\alpha_1}}} \times \cdots \times A/I_{J_{x^{\alpha_m}}}$ as the $A$-module representation of 
$A[x_{1},\ldots,x_{n}]/\mathfrak{a}$ w.r.t. $G$ (or w.r.t. $\prec$).  
If $I_{J_{x^{\alpha_i}}} = \{0\} $, we have $C_{J_{x^{\alpha_i}}} = A$,
$\text{for all } i = 1,\ldots,m$.  
This implies  $A[x_{1},\ldots,x_{n}]/\mathfrak{a} \cong A^m$, i.e. $A[x_{1},\ldots,x_{n}]/\mathfrak{a}$ has an $A$-module basis and it is free. 
We say that $A[x_{1},\ldots,x_{n}]/\mathfrak{a} $ has a free $A$-module representation w.r.t. $G$ (or w.r.t. $\prec$). 
Note that every basis of a finitely generated free $A$-module is finite.

We give below the definition of standard monomials in $A[x_{1},\ldots,x_{n}]$, w.r.t. an ideal $\mathfrak{a} \subseteq  A[x_1,\ldots,x_n]$. 
\begin{definition}
Let $\mathfrak{a} \subseteq A[x_1,\ldots,x_n]$ be an ideal.  We call a monomial $x^\alpha$ in $A[x_1,\ldots,x_n]$, a standard monomial w.r.t.
$\mathfrak{a}$ if none of the leading terms of the ideal divide the monomial, 
i.e. $x^\alpha \notin \langle \mathrm{lt}(\mathfrak{a})\rangle$. 
\end{definition}
We present below two results that give a necessary and sufficient condition for a finitely generated $A[x_{1},\ldots,x_{n}]/\mathfrak{a}$ to have a free $A$-module representation w.r.t. a monomial order. 
\begin{theorem}\label{Theorem Monic implies free}
Let $\mathfrak{a} \subseteq A[x_1,\ldots,x_n]$ be a non-zero ideal. Let $G$ be a  Gr\"obner basis for $\mathfrak{a}$ w.r.t. a monomial ordering, $\prec$.
If $G$ is monic then $A[x_1,\ldots,x_n]/\mathfrak{a}$ has a free $A$-module representation w.r.t. $G$.
\end{theorem}
\proof
Let $G = \{g_i : i = 1,\ldots,t\}$ be a monic Gr\"obner basis of the ideal.  
For a monomial $x^{\alpha}$, $J_{x^{\alpha}} =\{i : g_i\in G, \mathrm{lm}(g_i) \mid {x^{\alpha}}\}$. 
For a monomial $x^{\alpha}$ such that $\mathrm{lm}(g_i) \nmid x^{\alpha}$, for all  $g_i \in G$, we have $J_{x^{\alpha}} = \phi $.
Therefore, the leading coefficient ideal corresponding to those $x^{\alpha}$s, $ I_{J_{x^{\alpha}}}$ is  $\{0\}$ and the set of coset representatives 
$C_{J{_{x^\alpha}}}$ for $A/I_{{J_{{x^{\alpha}}}}}$ is the entire ring, $A$.
For a monomial $x^{\alpha}$ such that for some $g_i$, $\mathrm{lm}(g_i) \mid x^{\alpha}$, we have $J_{x^{\alpha}} \ne \phi$.
Since all the $g_i\in G$ are monic,  $ I_{{J_{x^{\alpha}}}} = \{1\}$, and therefore the set of coset representatives consist of only $0$. 
The only monomials that are part of the generating set therefore are monomials $x^\alpha$ such that $\mathrm{lm}(g_i) \nmid x^\alpha$,  
for all $g_i \in G$. Let  $S = \{x^\alpha + \mathfrak{a} : \mathrm{lm}(g_i) \nmid x^\alpha, \forall g_i \in G\}$.  Let $S'$ be any subset of $S$. Consider,
\begin{displaymath}
\sum_{x^{\alpha_j} + \mathfrak{a} \in S'} b_j( x^{\alpha_j} + \mathfrak{a}) = 0, \hspace{10pt} b_j \in A, \hspace{10pt} b_j \ne 0.
\end{displaymath} 
This implies,
\begin{displaymath}
 \sum_{x^{\alpha_j} + \mathfrak{a} \in S'} b_jx^{\alpha_j} + \mathfrak{a} = 0.
\end{displaymath}
Therefore we have,
 \begin{displaymath}
 \sum_{x^{\alpha_j} + \mathfrak{a} \in S'} b_j x^{\alpha_j} \in \mathfrak{a}.
 \end{displaymath} But that means  $\mathrm{lt}(g_i) \mid x^{\alpha_j} \text{ for some } j \text{ and for some } g_i\in G$, which is a contradiction.
 Therefore, $S$ is a basis for $A[x_{1},\ldots,x_{n}]/\langle G \rangle$. Thus the $A$-module,  $A[x_{1},\ldots,x_{n}]/\mathfrak{a}$ is free.
\endproof
Note that in the above theorem $A[x_1,\ldots,x_n]/\mathfrak{a}$ need not be finitely generated. If $A[x_1,\ldots,x_n]/\mathfrak{a}$ is finitely generated and 
the Gr\"obner basis of $\mathfrak{a}$ is monic, then there exists a $N \in \mathbb{N}$ such that $A[x_1,\ldots,x_n]/\mathfrak{a} \cong A^N$.

For the necessary condition we need the concept of short reduced Gr\"obner basis that we introduce in this paper.
\begin{definition}
Let $\mathfrak{a} \subseteq A[x_1,\ldots,x_n]$ be an ideal.  A reduced Gr\"obner basis $G$ of $\mathfrak{a}$  is called a short reduced Gr\"obner basis if 
for each $x^\alpha \in \mathrm{lm}(G)$, the length of the generating set for its leading coefficient ideal, $\mathrm{Gen} (\alpha, \mathfrak{a})$ is minimal. 
\end{definition}
We prove a lemma below that leads us to the necessary condition.
\begin{lemma} \label{lemma}
Let $\mathfrak{a} \subseteq A[x_1,\ldots,x_n]$ be a non-zero ideal such that $A[x_1,\ldots,x_n]/\mathfrak{a}$ is a finitely generated $A$-module and 
let $G$ be a short reduced Gr\"obner basis for $\mathfrak{a}$.
All the leading coefficient ideals associated with $G$ are either trivial or the entire ring $A$, if and only if $G$ is monic. 
\end{lemma}
\proof
Let $G = \{g_i : i = 1,\ldots,t\}$ be a short reduced Gr\"obner basis of the ideal, $\mathfrak{a}$.  Let the leading coefficient ideals associated with $G$, $I_{J_{x^{\alpha}}} $ 
be either trivial or $\langle 1 \rangle$. 
Suppose $G$ is not monic. Choose
$g \in G$ such that $g$ is not monic and for all $g_j \in G$ such that $\mathrm{lm}(g) \ne \mathrm{lm}(g_j)$, we have $\mathrm{lm}(g_j) \nmid \mathrm{lm}(g)$.
This is assured since if $g_j\in G$ and $\mathrm{lm}(g_j) \mid \mathrm{lm}(g)$ then either $\mathrm{lc}(g_j) = 1$ or $\mathrm{lc}(g_j) \ne 1$. If $\mathrm{lc}(g_j) = 1$ then 
$g$ can be removed from the reduced basis, this contradicts the uniqueness of the reduced Gr\"obner basis. If $g_j$ is not monic then we can choose $g_j$ as $g$.  
Let $\mathrm{lm}(g) = r$.  By assumption,
$I_{J_r} =  \langle 1 \rangle$ or $\{ 0 \}$. Since $J_r \ne \phi$, we have $I_{J_r} \ne  \{ 0 \}$. This implies, $\langle \mathrm{lc} (g_i) : g_i \in G, g_i \mid r \rangle = \langle 1 \rangle$. 
By choice of $g$ the ideal consists of only the leading coefficients of generators $g_i$ such that $\mathrm{lm}(g_i) = r$.  
Thus, we have the leading coefficient ideal of generators with the same degree as $g$, $\langle\mathrm{lc}(\mathrm{deg}(g),\mathfrak{a})\rangle = \langle 1 \rangle$. 
The generating set of minimal length for $\langle\mathrm{lc}(\mathrm{deg}(g),\mathfrak{a})\rangle$, $\mathrm{Gen} (\langle\mathrm{lc}(\mathrm{deg}(g),\mathfrak{a})\rangle)= \{1\}$. 
This implies $\mathrm{Gen} (\mathrm{deg}(g),\mathfrak{a}) = \{1\} $, since any other set of elements from $A$ that generate $\langle 1\rangle$ is of size strictly greater than $1$ and its length is not minimal.
To construct the  short reduced Gr\"obner basis we assumed that  $\mathrm{Gen} (\alpha,\mathfrak{a})$  is a generating set of minimal length. 
Therefore, $\mathrm{Gen} (\mathrm{deg}(g),\mathfrak{a}) = \{1\}$ and $g$ is a monic polynomial, which contradicts the fact that the basis $G$ is not monic.  
\par To prove the other direction, suppose $G$ is monic. For a monomial $x^{\alpha}$, we have $J_{x^{\alpha}} = \{i : g_i \in G, \mathrm{lm}(g_i) \mid {x^{\alpha}}\}$. 
For a monomial $x^{\alpha}$ such that $\mathrm{lm}(g_i) \nmid x^{\alpha}$, for all $g_i \in G$, we have  $J_{x^{\alpha}} = \phi$. 
This implies that the leading coefficient ideal,  $I_{J_{x^{\alpha}}} = \{0\}$.
 For a monomial $x^{\alpha}$ such that  $\mathrm{lm}(g_i) \mid x^{\alpha}$ for some $i \in \{1,\ldots, t\}$, we have $ J_{x^{\alpha}} \ne \phi$ and  $I_{J_{x^{\alpha}}} = \langle \mathrm{lc}(g_i) : i \in J_{x^{\alpha}}\rangle $.
Since  $G$ is monic this implies, $I_{J_{x^{\alpha}}}=\langle 1\rangle$.
\endproof
We are now ready to give the necessary condition of the characterization.
\begin{theorem}
Let $\mathfrak{a} \subseteq A[x_1,\ldots,x_n]$ be a non-zero ideal such that $A[x_1,\ldots,x_n]/\mathfrak{a}$ is a finitely generated $A$-module and 
let $G$ be a short reduced Gr\"obner basis for $\mathfrak{a}$.  If $A[x_1,\ldots,x_n]/\mathfrak{a}$ has a free $A$-module representation w.r.t. $G$, then $G$ is monic. 
\end{theorem}
\proof
Let $G = \{g_i : i = 1,\ldots,t\}$ be a short reduced Gr\"obner basis of the ideal, $\mathfrak{a}$. 
Since  $A[x_1,\ldots,x_n]/\mathfrak{a}$ has a free $A$-module representation w.r.t. $G$, there are only two possibilities for the  leading coefficient ideals associated with $G$, 
$I_{J_{x^{\alpha}}} = \{0\}$ or $I_{J_{x^{\alpha}}} =\langle 1 \rangle$.
From Lemma ~\ref{lemma}, $G$ is a monic basis. 
\endproof

We state the characterization result as follows. 
\begin{proposition}\label{Proposition for characterization}
Let $\mathfrak{a} \subseteq A[x_1,\ldots,x_n]$ be a non-zero ideal such that $A[x_1,\ldots,x_n]/\mathfrak{a}$ is finitely generated. 
Let $G$ be a  short reduced Gr\"obner basis for $\mathfrak{a}$ w.r.t. some monomial ordering, $\prec$. Then, $A[x_1,\ldots,x_n]/\mathfrak{a}$ has a free $A$-module representation w.r.t. $G$,
if and only if $G$ is monic.
\end{proposition}
The necessity for $\mathrm{Gen} (\alpha,\mathfrak{a})$ to be of minimal length  can be illustrated by the following example. 
\begin{example} Consider Example ~\ref{Example}. We have $\mathfrak{a}\subseteq \mathbb{Z}[x,y]$ given by its Gr\"obner basis, $G= \{ 3 x^2, 5x^2,y\}$. 
The Gr\"obner basis $G= \{ 3 x^2, 5x^2,y\}$ is a reduced Gr\"obner basis for the ideal when we select the generators for the leading coefficient ideal for 
each leading monomial in $G$, $\mathrm{Gen} (\langle\mathrm{lc}(\alpha,\mathfrak{a})\rangle)$ as the same set given in the example. It is not a monic basis. 
But one can see that $\mathbb{Z}[x,y]/\langle G \rangle$ is free.
A  short reduced Gr\"obner basis for the same ideal $\mathfrak{a}$, determined by considering the $\mathrm{gcd}$ of the generators of the leading coefficient ideal of each monomial in $G$, is $\{x^2,y\}$.
The generating set $\mathrm{Gen} (\alpha,\mathfrak{a})$ is of minimal length when the  $\mathrm{gcd}$ of the generators is considered.
The   short reduced Gr\"obner basis is monic and leads us to the correct conclusion that $\mathbb{Z}[x,y]/\langle G \rangle$ is free.
\end{example}
\section{Macaulay-Buchberger Basis Theorem Over Rings}
\label{M-B}
\noindent The Macaulay Basis Theorem  can be extended directly from fields to rings, i.e. $S = \{ x ^\alpha + \mathfrak{a} : x^\alpha \hspace{5pt}\text{is a standard monomial, i.e. } 
x^\alpha \notin \langle \mathrm{lt}(\mathfrak{a})\rangle\}$ is an $A$-module basis for  $A[x_1,\ldots,x_n]/\mathfrak{a}$ if $A[x_1,\ldots,x_n]/\mathfrak{a}$ is free.
We extend below the Macaulay-Buchberger Basis Theorem over rings.
\begin{theorem}[Macaulay-Buchberger Basis Theorem Over Rings]\label{Macaulay Buchberger}
 Let $G = \{g_1, \ldots, g_t\}$ be a short reduced Gr\"obner basis for an ideal $\mathfrak{a} \subseteq A[x_1,\ldots,x_n]$. 
Suppose $G$ is monic then  an $A$-module basis for $A[x_1,\ldots,x_n]/\langle \mathfrak{a} \rangle$ is given by $S = \{ x ^\alpha + \mathfrak{a} : \mathrm{lm}(g_i)\nmid x^\alpha, i=1,\ldots,t\}$.
\end{theorem}
\proof
From the characterization result we have that $A[x_1,\ldots,x_n]/\mathfrak{a}$ has a free $A$-module representation w.r.t. $G$ if and only if the short reduced Gr\"obner basis for $\mathfrak{a}$ is monic. 
The proof is along the same lines as Theorem ~\ref{Theorem Monic implies free}.    
\endproof
In the above theorem the necessity for $G$ to be a short reduced Gr\"obner basis can be explained as follows.
Unlike in the case of fields, for any Gr\"obner basis $G$ in $A[x_1,\ldots,x_n]$  and any $x^\alpha \in {\mathbb{Z}}_{\ge 0}^n$, $x^\alpha \in \langle \mathrm{lt}(\mathfrak{a})\rangle$ 
does not imply  $\mathrm{lt}(g_i)\mid x^\alpha$, for some $g_i \in G $. 
If $G$ is the short reduced Gr\"obner basis and  $A[x_1,\ldots,x_n]/\mathfrak{a}$ has a free $A$-module representation w.r.t. $G$ 
then each $g_i \in G$ is monic. We then have $x^\alpha \in \langle \mathrm{lt}(\mathfrak{a})\rangle$ if and only if  $\mathrm{lt}(g_i)\mid x^\alpha$, for some $g_i \in G$.

This is explained in the following example.
\begin{example}
 Consider an ideal $\mathfrak{a} \subseteq \mathbb{Z}[x,y]$ generated by the Gr\"obner basis, $G= \{3x^2, 5x^2,y\}$. It can be seen that  $3x^2$ cannot be reduced further by 
 $\{5x^2,y\} $. Similarly, $5x^2$ is minimal w.r.t. the set  $\{3x^2,y\} $ and $y$ is minimal w.r.t.  $\{3x^2, 5x^2\}$. Therefore $G$ is a minimal Gr\"obner basis. 
Consider the monomial $x^2$. It is in the ideal,  $\langle \mathrm{lt}(\mathfrak{a})\rangle$ since $x^2 = (2)(5)x^2 - (3)(3)x^2$. But none of the leading terms divide $x^2$,
 $5x^2 \nmid x^2$, $3x^2 \nmid x^2$ and  $y \nmid x^2$.  Now consider a short reduced Gr\"obner basis  of the ideal, $\{x^2,y\}$. A $\mathbb{Z}$-module basis of $\mathbb{Z}[x,y]/\mathfrak{a}$ 
 is the set of residue classes of  $\{ x^\alpha : x^2 \nmid x^\alpha, y \nmid x^\alpha \} = \{1  + \mathfrak{a}, x + \mathfrak{a}\}$.
 \end{example}
 
In the zero-dimensional case, Gr\"obner basis generalizes the notion of Gaussian elimination by generating a maximum possible triangular system of polynomial equations over a field 
\cite[Theorem 2.2.7]{Adams:1994:introtogrobnerbasis}. We can extend this result to the case of polynomials over rings too, as shown below.
 \begin{theorem}
Let $\mathfrak{a}$ be an ideal in $A[x_1,\ldots,x_n]$ and let $G = \{g_1,\ldots, g_t\}$ be a monic short reduced Gr\"obner basis for $\mathfrak{a}$ w.r.t. a monomial order, $\prec$.
We have from the characterization that $A[x_1,\ldots,x_n]/\mathfrak{a}$ has a free $A$-module representation w.r.t. $G$. The following statements are equivalent. 
\begin{enumerate}[(i)]
\item For each $i = 1,\ldots, n$, there exists $j\in \{1,\ldots, t\}$ such that $\mathrm{lm}(g_j) = x_i ^\nu$ for some $\nu \in \mathbb{N}$.
\item The rank of the free $A$-module, $A[x_1,\ldots,x_n]/\mathfrak{a}$ is finite. 
\end{enumerate}
 \end{theorem}
 \proof
 (i)$\Longrightarrow$ (ii). The short reduced Gr\"obner basis of the ideal $\mathfrak{a}$ is monic implies that the $A$-module $A[x_1,\ldots,x_n]/\mathfrak{a}$ is free and the
 basis is the set of cosets of monomials such that 
none of the leading monomials of the Gr\"obner basis divide the monomial.  Since for every $i = 1,\ldots, n$, there exists $j\in \{1,\ldots, t\}$ 
such that $\mathrm{lm}(g_j) = x_i ^\nu$ for some $\nu \in \mathbb{N}$, there are only finitely many power products which are reduced w.r.t. $G$ and hence the dimension of the free $A$-module is finite.
 
(ii)$\Longrightarrow$ (i).  We have that the rank of the free $A$-module $A[x_1,\ldots,x_n]/\mathfrak{a}$ is finite. Assume for some $i \in \{1,\ldots,n\}$ there is no $j \in \{1,\ldots,t\}$ 
such that $\mathrm{lm}(g_j) =  x_i ^\nu$ for some $\nu \in \mathbb{N}$. Then the powers of $x_i$, $1,x_i,x_i^2,\ldots$ are linearly independent and therefore 
contradicts the finite rank of the $A$-module, $A[x_1,\ldots,x_n]/\mathfrak{a}$.
 \endproof
Our characterization (Proposition ~\ref{Proposition for characterization}) and the Macaulay-Buchberger basis theorem (Theorem ~\ref{Macaulay Buchberger}) give rise to 
an algorithm (Algorithm 1) to compute an $A$-module basis of a 
free residue class ring, $A[x_1,\ldots,x_n]/\mathfrak{a}$, when it is finitely generated. 
The correctness of the algorithm directly follows from the characterization of a free $A[x_1,\ldots,x_n]/\mathfrak{a}$ and the Macaulay-Buchberger basis theorem. 
The termination of the algorithm is ensured since we have a finitely generated and free $A$-module, $A[x_1,\ldots,x_n]/\mathfrak{a}$, which implies it has a finite basis.
\begin{algorithm}\label{Algorithm}
\caption{Finding the A-module basis of finitely generated residue class  polynomial rings over rings} 
\begin{algorithmic}
\STATE \textbf{Input} $A[x_1,\ldots,x_n]/\mathfrak{a}$, \\
$G=\{g_1,\ldots,g_t\}$, a short reduced Gr\"obner basis of $\mathfrak{a}$, \\
$S \subseteq A[x_1,\ldots,x_n]/\mathfrak{a}
$,\\
$M\subseteq {\mathbb{Z}}_{\ge 0}^n$.
\STATE \textbf{Output} $S = A$-module basis of $A[x_1,\ldots,x_n]/\mathfrak{a}$.
\STATE $S = \phi$, $M=\phi$
\IF {$G$ is not monic}
\STATE $A$-module basis for $A[x_1,\ldots,x_n]/\mathfrak{a}$ does not exist.
\ELSE
\WHILE {${\mathbb{Z}}_{\ge 0}^n \setminus M \neq \phi$}
\STATE for each monomial $x^\alpha \in {{\mathbb{Z}}_{\ge 0}^n \setminus M}$
\IF {$\mathrm{lm}(g_i) \mid x^\alpha$   for some  $i \in \{1,\ldots, t\}$}
\STATE $M= M \cup \{x^\beta \in {\mathbb{Z}}_{\ge 0}^n : x^\alpha \mid  x^\beta\}$
\ELSE 
\STATE $S =S \cup \{ x^\alpha + \mathfrak{a} \}$ 
\STATE $M = M \cup \{x^\alpha\}$
\ENDIF
\ENDWHILE
\ENDIF
\end{algorithmic}
\end{algorithm}
 \section{Special cases, $A = \mathbb{Z}$ and $A = \Bbbk[\theta_1,\ldots,\theta_m]$}
\label{specialcases}
\noindent Now we look at two special cases $A = \mathbb{Z}$ and $A = \Bbbk[\theta_1,\ldots,\theta_m]$. 
In the case of $A= \mathbb{Z}$ we make use of the fact that the ring is a PID in the proof and in the case of 
$A = \Bbbk[\theta_1,\ldots,\theta_m]$ we rely on the existence of a unique strong reduced Gr\"obner basis for any ideal in $\Bbbk[\theta_1,\ldots,\theta_m]$ \citep{Nabeshima:2009:polyringoverpolyring}. 
\subsection{Special case : $A = \mathbb{Z}$ }
For general rings, we arrived at the conclusion that if a finitely generated $A$-module, $A[x_1,\ldots,x_n]/\mathfrak{a}$, has a free $A$-module representation w.r.t. a monomial order, $\prec$ then 
its short reduced Gr\"obner basis $G$ w.r.t. $\prec$ is monic, by finding a contradiction to the minimality of the length of the generating set. 
But here we argue that if the reduced Gr\"obner basis is not monic then there exists some $g$ such that $\langle\mathrm{lc}(\mathrm{deg}(g),\mathfrak{a})\rangle = \langle c \rangle$, 
where $c \in \mathbb{Z}$ and $c \ne 1$. 
But this means that the set of coset representatives for the monomial $\mathrm{lm}(g)$, $C_{J_{\mathrm{lm}(g)}}$ can never be zero. 
This is a contradiction to our assumption that $\mathbb{Z}[x_1\ldots,x_n]/\mathfrak{a} \cong \mathbb{Z}^N$ w.r.t. $G$. 

An example which illustrates the characterization is the case when the ideal in $\mathbb{Z}[x_1,\ldots, x_n]$ is a lattice ideal. 
A lattice ideal, $\mathfrak{a}_\mathcal{L}$ in $\Bbbk[x_1,\ldots,x_n]$ is defined as the binomial ideal generated by $\{x^{v^+} - x^{v^-}\}$ where $v^+$ and $v^-$ are non-negative with disjoint support
and $v^+ - v^- \in \mathcal{L}$, where $\mathcal{L}$
is a lattice. Lattice ideals in polynomial rings over $\mathbb{Z}$ can be defined in the same way. 
In this case, the binomial ideal is generated over the polynomial ring, $\mathbb{Z}[x_1,\ldots,x_n]$. 
The generators of the ideal are binomials with the terms having opposite sign and the coefficients of both the terms equal to absolute value $1$. 
When we compute the Gr\"obner basis of the ideal, at every stage of the computation -- $\mathrm{S}$-polynomial calculation and reduction  --  
we add generators that are binomials with terms having opposite sign and coefficients of absolute value $1$. Therefore the Gr\"obner basis is monic
which implies that the short reduced Gr\"obner basis of the ideal is also monic.
From the characterization we have that the quotient ring $\mathbb{Z}[x_1,\ldots,x_n]/\mathfrak{a}_\mathcal{L}$ is free. We will now formally state the result.
\begin{theorem}
The quotient ring $\mathbb{Z}[x_1,\ldots,x_n]/\mathfrak{a}_\mathcal{L}$, where $\mathfrak{a}_\mathcal{L}$ is a lattice ideal in $\mathbb{Z}[x_1,\ldots,x_n]$, is free. 
\end{theorem}
Another application of the characterization is that we can identify ideals in  $\mathbb{Z}[x_1,\ldots,x_n]$ for which all 
the ideals in the corresponding residue class polynomial rings are integer lattices, i.e. all ideals are ideal lattices.  We formally state that below.
\begin{theorem}
Let $\mathfrak{a}\subseteq\mathbb{Z}[x_1,\ldots,x_n]$ be an ideal. All ideals in $\mathbb{Z}[x_1,\ldots,x_n]/\mathfrak{a}$ are integer lattices if and only if the 
short reduced Gr\"obner basis of $\mathfrak{a}$ is monic for some monomial order, $\prec$.
\end{theorem}
\subsection{Special case : $A = \Bbbk[\theta_1,\ldots,\theta_m]$}
We now consider the next special case where the coefficient ring is itself a polynomial ring over the field $\Bbbk$, $\Bbbk[\theta_1,\ldots,\theta_m]$. 
We consider here the definition of strong reduced Gr\"obner basis defined in \citep{Nabeshima:2009:polyringoverpolyring} and give an 
alternate characterization for this definition as well. The strong reduced Gr\"obner basis defined in  \citep{Nabeshima:2009:polyringoverpolyring} is specific to polynomial rings over polynomial rings.
\par For ease of notation, we denote the indeterminates $\theta_1,\ldots,\theta_m$ as $\Theta$ and $x_{1},\ldots,x_{n}$ as $X$. 
In the polynomial ring $\Bbbk[\theta_1,\ldots,\theta_m][x_{1},\ldots,x_{n}]$, we define leading monomial, leading coefficient and 
leading term w.r.t. the monomial ordering on $X$ indeterminates and we denote them as $\mathrm{lm}_X$, $\mathrm{lc}_X$ and $\mathrm{lt}_X$.
\begin{definition}[Strong Reduced Gr\"obner Basis\citep{Nabeshima:2009:polyringoverpolyring}] \label{Nabeshima GB}Let $\prec_{X,\Theta} := (\prec_1, \prec_2)$ be a block ordering,
$\mathfrak{a}$ an ideal in $\Bbbk[\theta_1,\ldots,\theta_m][x_{1},\ldots,x_{n}]$ and $G$ a subset of $\Bbbk[\theta_1,\ldots,\theta_m][x_{1},\ldots,x_{n}]$. 
For $e \in \mathrm{lm}_{X}(G)$, let $G_e = \{f \in G \mid \mathrm{lm}_{X}(f) = e\}$. Then a strong reduced Gr\"obner basis $G$ for $\mathfrak{a}$ w.r.t.  $\prec_1$ and $\prec_2$ is a 
Gr\"obner basis for $\mathfrak{a}$ in $\Bbbk[\theta_1,\ldots,\theta_m][x_{1},\ldots,x_{n}]$ such that for all $p \in G$, 
\begin{enumerate}[(i)]
\item no term in $p$ lies in $\langle \mathrm{lt} (G \backslash \{p\})\rangle$ in $\Bbbk[\theta_1,\ldots,\theta_m, x_{1},\ldots,x_{n}]$ w.r.t. $\prec_{X,\Theta}$,
\item no term in $p$ lies in $\langle \mathrm{lt}_{X} (G \backslash \{p\})\rangle$ in $\Bbbk[\theta_1,\ldots,\theta_m][x_{1},\ldots,x_{n}]$ w.r.t. $\prec_1$,
\item for $e \in \mathrm{lm}_{X}(G)$, $\mathrm{lc}_{X}(G_e)$ is the reduced Gr\"obner basis for an ideal generated by itself w.r.t. $\prec_2$ in the quotient 
ring $\Bbbk[\theta_1,\ldots,\theta_m]/\mathfrak{J}_e$ where $\mathfrak{J}_e$ is an ideal generated by 
$\mathfrak{L} = \{ \mathrm{lc}_{X}(g) \mid g \in G \backslash G_e \hspace{5pt}\text{such that} \hspace{5pt}\mathrm{lm}(g)\mid e\}$.
\end{enumerate}
\end{definition}
We give below the necessary and sufficient condition for a finitely generated $\Bbbk[\theta_1,\ldots,\theta_m][x_{1},\ldots,x_{n}]/\mathfrak{a}$ to be a 
free $\Bbbk[\theta_1,\ldots,\theta_m]$-module, in terms of strong reduced Gr\"obner basis.
\begin{theorem}
Let $\mathfrak{a} \subseteq \Bbbk[\theta_1,\ldots,\theta_m][x_{1},\ldots,x_{n}]$ be a non-zero ideal such that $\Bbbk[\theta_1,\ldots,\theta_m][x_{1},\ldots,x_{n}]/\mathfrak{a}$ is finitely generated. 
Let $\prec_{X,\Theta} := (\prec_1, \prec_2)$ be a block ordering and let $G$ be the strong  reduced Gr\"obner basis for $\mathfrak{a}$ w.r.t.  $\prec_1$ and $\prec_2$ . Then,
$\Bbbk[\theta_1,\ldots,\theta_m][x_1,\ldots,x_n]/\mathfrak{a}$ has a free $\Bbbk[\theta_1,\ldots,\theta_m]$-module representation w.r.t. $G$
if and only if $G$ is monic.
\end{theorem}
\proof
The proof for if $G = \{g_i : i = 1,\ldots,t\}$ is a monic Gr\"obner basis of the ideal, $\mathfrak{a}$ w.r.t.  $\prec_1$ and $\prec_2$ then
$\Bbbk[\theta_1,\ldots,\theta_m][x_{1},\ldots,x_{n}]/\mathfrak{a}$ has a free $\Bbbk[\theta_1,\ldots,\theta_m]$-module representation w.r.t. $G$ is  same as Theorem ~\ref{Theorem Monic implies free},
with $A = \Bbbk[\theta_1,\ldots,\theta_m]$.
The set, $S = \{x^\alpha + \mathfrak{a} : \mathrm{lm}(g_i) \nmid x^\alpha, \forall g_i \in G\}$ forms the $A$-module basis. 
Note that in the proof, $\Bbbk[\theta_1,\ldots,\theta_m][x_{1},\ldots,x_{n}]/\mathfrak{a}$ need not be finitely generated. 
If $\Bbbk[\theta_1,\ldots,\theta_m][x_{1},\ldots,x_{n}]/\mathfrak{a}$ is finitely generated, then there exists a $N \in \mathbb{N}$ such that
$\Bbbk[\theta_1,\ldots,\theta_m][x_{1},\ldots,x_{n}]/\mathfrak{a} \cong \Bbbk[\theta_1,\ldots,\theta_m]^N$.

Conversely, let $\mathfrak{a} \subseteq \Bbbk[\theta_1,\ldots,\theta_m][x_{1},\ldots,x_{n}]$ be an ideal such that
$\Bbbk[\theta_1,\ldots,\theta_m][x_{1},\ldots,x_{n}]/\mathfrak{a}$ is a finitely generated $\Bbbk[\theta_1,\ldots,\theta_m]$-module and 
let $G$ be the strong reduced Gr\"obner basis for $\mathfrak{a}$. \\Assume $\Bbbk[\theta_1,\ldots,\theta_m][x_{1},\ldots,x_{n}]/\mathfrak{a}$ has a free $\Bbbk[\theta_1,\ldots,\theta_m]$-module representation 
w.r.t. $G$,
we have to prove that $G$ is monic.
In Definition ~\ref{Nabeshima GB} of strong reduced Gr\"obner basis, consider the third condition. 
We have, for any monomial $x^\alpha$ in the strong reduced Gr\"obner basis,  $\mathrm{lc}_{X}(G_{x^\alpha})$ is the reduced Gr\"obner basis in the ring 
$\Bbbk[\theta_1,\ldots,\theta_m]/\mathfrak{J}_{x^\alpha}$. 
Suppose the strong reduced Gr\"obner basis $G$ is not monic. 
Then, there exists a $g\in G$ such that $g$ is not monic 
w.r.t. $\prec_1$ and $\mathrm{lm}(g_j) \nmid \mathrm{lm}(g)$, for all  $g_j \in G$, such that  $\mathrm{lm}(g_j) \ne \mathrm{lm}(g) $. 
Let $\mathrm{lm}_{X}(g) = r$. Since $g$ is not monic,  $\langle \mathrm{lc}_{X}(G_r)\rangle \ne 1$ in $\Bbbk[\theta_1,\ldots,\theta_m]/\mathfrak{J}_r$. 
But $\Bbbk[\theta_1,\ldots,\theta_m][x_{1},\ldots,x_{n}]/\langle G \rangle \cong {\Bbbk[\theta_1,\ldots,\theta_m]}^N, \hspace{5pt} N \in \mathbb{N}$, implies that $I_{J_r} = \langle 1 \rangle$. 
This means $\langle\mathrm{lc}_{X}(g_i) : g_i \in G, \mathrm{lm}(g_i)\mid r\rangle = \langle 1 \rangle$. 
But $\langle\mathrm{lc}_{X}(g_i) : g_i \in G, \mathrm{lm}(g_i)\mid r\rangle =\langle\mathrm{lc}_{X} (G_r)\rangle + \mathfrak{J}_r$. 
We have $\langle\mathrm{lc}_{X} (G_r)\rangle + \mathfrak{J}_r = \langle 1 \rangle$. 
This means $\langle\mathrm{lc}_{X} (G_r)\rangle = \langle 1 \rangle$ in $\Bbbk[\theta_1,\ldots,\theta_m]/\mathfrak{J}_{r}$ 
which contradicts the fact that $G$ is not monic.
Therefore $G$ is a monic basis.
\endproof
\begin{example}
 Consider the ring $\Bbbk[a][x]$. We have a Gr\"obner basis $G = \{f_1,f_2\}$ where $f_1 = a^2x-a, f_2 = (a^3 - 1)x - a^2 + 1$. The set of all leading monomials in $G$ is $\{x\}$.
 We have $G_x = \{f_1, f_2\}$ 
and $\mathrm{lc}_{\{x\}} (G_x) = \{a^2, a^3-1\}$. Since there are no other monomials other than $x$ we have $\mathfrak{L} = \phi$, which implies $\Bbbk[a]/\mathfrak{J}_x = \Bbbk[a]$. 
So now as per the third condition in the 
definition we have that $\mathrm{lc}_{\{x\}} (G_x)$ should be a reduced Gr\"obner basis in $\Bbbk[a]$. The reduced Gr\"obner basis of $\{a^2, a^3-1\} = \{1\}$. 
Therefore, we compute a new polynomial $g$ such that $\langle g \rangle = \langle G \rangle$, $\langle \mathrm{lm}_{\{x\}}(g) \rangle = \langle \mathrm{lm}_{\{x\}}(G)\rangle$ 
and $\mathrm{lc}_{\{x\}} (g) = \{1\}$. 
This $g = af_1 - f_2 = x-1$. Thus $\{g\}$ is the strong reduced Gr\"obner basis and it is monic. This implies $\Bbbk[a][x]/\langle G \rangle$ is a free $\Bbbk[a]$--module. 
\end{example}
Now, we proceed to study how strong reduced Gr\"obner bases is related to short reduced Gr\"obner bases.
 Let $\prec_{X,\Theta} := (\prec_1, \prec_2)$ be a block ordering and $\mathfrak{a}$ an ideal in $\Bbbk[\theta_1,\ldots,\theta_m][x_{1},\ldots,x_{n}]$ .
 For each $x^\alpha \in \mathfrak{a}$, we can construct a  generating set of minimal length for $\mathrm{Gen}(\alpha, \mathfrak{a})$  
 by taking the reduced Gr\"obner basis w.r.t. $\prec_2$ as the set of generators for any ideal $I$ in $\Bbbk[\theta_1,\ldots,\theta_m]$ and for any polynomial $h \in \Bbbk[\theta_1,\ldots,\theta_m]$,
by  considering the mapping $\eta_I(h)$ as the normal form of $h$ w.r.t. $I$ and $\prec_2$. 
The corresponding Pauer's reduced Gr\"obner basis is the short reduced Gr\"obner basis.  
One can see from the below result that short reduced Gr\"obner basis is also the strong reduced Gr\"obner basis.
\begin{proposition}
Let $\prec_{X,\Theta} := (\prec_1, \prec_2)$ be a block ordering and $\mathfrak{a} \subseteq \Bbbk[\theta_1,\ldots,\theta_m][x_{1},\ldots,x_{n}]$ be an ideal. 
Let $G$ be the short reduced Gr\"obner basis of $\mathfrak{a}$ constructed by taking the reduced Gr\"obner basis w.r.t. $\prec_2$ as the set of generators for any ideal 
$I$ in $\Bbbk[\theta_1,\ldots,\theta_m]$  and for any polynomial $h \in \Bbbk[\theta_1,\ldots,\theta_m]$,
by  considering the mapping $\eta_I(h)$ as the normal form of $h$ w.r.t. $I$ and $\prec_2$. Then $G$ is the strong reduced Gr\"obner basis.
\proof
We have to prove that the short reduced Gr\"obner basis of $\mathfrak{a}$, $G = \{g_i : i = 1, \ldots, t\}$ satisfies that the three conditions mentioned in Definition ~\ref{Nabeshima GB}. 
\begin{enumerate}[(i)]
\item Suppose a term in $p \in G$ lies in  $\langle \mathrm{lt} (G \backslash \{p\})\rangle$ in $\Bbbk[\theta_1,\ldots,\theta_m, x_{1},\ldots,x_{n}]$ w.r.t. $\prec_{X,\Theta}$. 
Let that term be $a\theta^\beta x^\alpha$, where $a\in \Bbbk$, $\theta^\beta$ is a monomial in $\theta_1,\ldots, \theta_m$, $\beta \in {\mathbb{Z}}_{\ge 0}^m$ and $\alpha \in {\mathbb{Z}}_{\ge 0}^n$. 
The coefficient of the term, $a\theta^\beta x^\alpha$ in $\Bbbk[\theta_1,\ldots,\theta_m][x_{1},\ldots,x_{n}]$ is $a\theta^\beta$.  We have,
 $\theta^\beta \in \langle \mathrm{lc}_{X}(\mathrm{lt}(G \backslash \{p\}))\rangle$. Since $\langle \mathrm{lc}_{X}(\mathrm{lt}(G \backslash \{p\}))\rangle$ is a monomial ideal,  
 $\theta^\beta$ is divisible by  $\mathrm{lc}_{X}(\mathrm{lt}(g_i))$ for some $g_i \in G \backslash \{p\}$.
Since we have a block ordering with ordering of $\{x_1,\ldots, x_n\}$ variables taking precedence over the $\{\theta_1,\ldots,\theta_m\}$ variables, it can be seen that 
$\mathrm{lc}_{X}(\mathrm{lt}(g_i))$ is the leading monomial of the polynomial $\mathrm{lc}_{X}(g_i)$ in $\Bbbk[\theta_1,\ldots,\theta_m]$. Let us first look at the case when $a\theta^\beta$ is 
a term in the leading coefficient, $\mathrm{lc}_{X}(p)$. We have that $\mathrm{lc}_{X}(p)$ is an element of the reduced Gr\"obner basis of the ideal 
$\langle \mathrm{lc}_{X}(\mathrm{deg}(\mathrm{lc}_{X}(p)), \mathfrak{a} )\rangle$ and therefore no term in $\mathrm{lc}_{X}(p)$ (including $a\theta^\beta$) can be reduced 
by the leading monomials of the polynomials in the reduced Gr\"obner basis other than itself. We therefore have a contradiction. Now we consider the case when $a\theta^\beta$ is not a term in the 
leading coefficient. Then we have from the definition of reduced Gr\"obner basis that $\theta^\beta$ is the normal form w.r.t. the ideal 
$\langle \mathrm{lc}(\mathrm{deg}(\theta^\beta),\mathfrak{a})\rangle$ and $\prec_2$. But since $\theta^\beta \in \langle \mathrm{lc}_{X}(\mathrm{lt}(G \backslash \{p\}))\rangle$,
we have that the normal form is zero which is a contradiction.  
\item  Suppose a term in $p \in G$ lies in  $\langle \mathrm{lt}_{X}(G \backslash \{p\})\rangle$ in $\Bbbk[\theta_1,\ldots,\theta_m][x_{1},\ldots,x_{n}]$ w.r.t.
$\prec_1$.  Let  $h(\theta)\in\Bbbk[\theta_1,\ldots,\theta_m]$ be the coefficient of that term in $\Bbbk[\theta_1,\ldots,\theta_m][x_{1},\ldots,x_{n}]$. We have,
 $h(\theta) \in \langle \mathrm{lc}_{X}(\mathrm{lt}_{X}(G \backslash \{p\}))\rangle$. Let us first look at the case when $h(\theta)$ is the leading coefficient of $p$, $\mathrm{lc}_{X}(p)$. 
 We have that $\mathrm{lc}_{X}(p)$ is an element of the reduced Gr\"obner basis of the ideal $\langle \mathrm{lc}_{X}(\mathrm{deg}(\mathrm{lc}_{X}(p)), \mathfrak{a} )\rangle$ and 
 therefore cannot be reduced by the leading monomials of the polynomials in the reduced Gr\"obner basis other than itself. We therefore have a contradiction. 
 Now we consider the case when $h(\theta)$ is not the leading coefficient. Then we have from the definition of reduced Gr\"obner basis that $h(\theta)$ is the normal form w.r.t. the ideal 
 $\langle \mathrm{lc}(\mathrm{deg}(h(\theta)),\mathfrak{a})\rangle$ and $\prec_2$. Since $h(\theta) \in \langle \mathrm{lc}_{X}(\mathrm{lt}(G \backslash \{p\}))\rangle$, we have 
 that the normal form is zero which is a contradiction.  
 \item In short reduced Gr\"obner basis, for each $e \in \mathrm{lm}_{X}(G)$  we choose the reduced Gr\"obner basis for the ideal,
 $\langle \mathrm{lc}_{X}(g_i) : g_i \in G, \mathrm{lm}(g_i) \mid e \rangle$ in $\Bbbk[\theta_1,\ldots,\theta_m]$ as its generators. 
 We have $\mathrm{lc}_{X}(G_e) = \{\mathrm{lc}_{X}(g_i) : g_i \in G, \mathrm{lm}_{X}(g_i) = e\}$. Since  $\mathrm{lc}_{X}(G_e)$ is a 
 subset of  $\mathrm{Gen}(\langle \mathrm{lc}_{X}(g_i) : g_i \in G, \mathrm{lm}(g_i) \mid e \rangle) $, it is the reduced Gr\"obner basis of the ideal
 generated by itself in $\Bbbk[\theta_1,\ldots,\theta_m]$. In the short reduced Gr\"obner basis $G$ over $\Bbbk[\theta_1,\ldots,\theta_m][x_{1},\ldots,x_{n}]$, the coefficient of 
 each term of a basis element is the normal form w.r.t. the ideal formed by the leading coefficients of all $g_i \in G$ such that $\mathrm{lm}(g_i)$ divides the term. 
 Therefore,  $\mathrm{lc}_{X}(G_e)$ is the reduced Gr\"obner basis of the ideal generated by itself in $\Bbbk[\theta_1,\ldots,\theta_m]/\mathfrak{J}_e$ where $\mathfrak{J}_e$ is an 
 ideal generated by $\mathfrak{L} = \{ \mathrm{lc}_{X}(g) \mid g \in G \backslash G_e \hspace{5pt}\text{such that} \hspace{5pt}\mathrm{lm}(g)\mid e\}$. 
\end{enumerate}
Thus the short Gr\"obner basis $G$ satisfies the three conditions of Definition ~\ref{Nabeshima GB}, hence it is the strong reduced Gr\"obner basis.
\endproof
\end{proposition}
\begin{proposition}
Let $\prec_{X,\Theta} := (\prec_1, \prec_2)$ be a block ordering and \\$\mathfrak{a} \subseteq \Bbbk[\theta_1,\ldots,\theta_m][x_{1},\ldots,x_{n}]$ be an ideal. 
Let $G$ be the strong reduced Gr\"obner basis for $\mathfrak{a}$ w.r.t. $\prec_{X,\Theta}$. Then $G$ is the short reduced Gr\"obner basis for $\mathfrak{a}$.
\end{proposition}
\proof
For every ideal $\mathfrak{a}$ in $\Bbbk[\theta_1,\ldots,\theta_m][x_{1},\ldots,x_{n}]$ there exists a short reduced Gr\"obner basis. 
We have shown in the above proposition that the short reduced Gr\"obner basis is a strong reduced Gr\"obner basis. 
We have that strong reduced Gr\"obner basis for an ideal $\mathfrak{a}$ in $\Bbbk[\theta_1,\ldots,\theta_m][x_{1},\ldots,x_{n}]$ is unique 
\citep{Nabeshima:2009:polyringoverpolyring}. This implies that the strong reduced Gr\"obner basis $G$ is the short reduced Gr\"obner basis.
\endproof
We would like to mention here that Nabeshima's strong reduced Gr\"obner basis should not be confused with the concept of strong Gr\"obner basis defined for ideals
in polynomial rings over $\mathrm{PID}$s \citep{Adams:1994:introtogrobnerbasis}. To avoid confusion, we mention the former as strong reduced Gr\"obner basis and 
the latter as strong Gr\"obner basis. We give below the definition of strong Gr\"obner basis.
\begin{definition}
Let $G= \{g_1, \ldots, g_t\}\subseteq A[x_1,\ldots, x_n]$ be a set of non-zero polynomials. $G$ is said to be a strong Gr\"obner basis for the ideal $\mathfrak{a}$ it generates,
if for each $f \in \mathfrak{a}$, there exists a $g_i \in G$ such that $\mathrm{lt}(g_i) \mid \mathrm{lt}(f)$.  
\end{definition}
The definition does not require $A$ to be a $\mathrm{PID}$ but it can be easily shown that strong Gr\"obner bases exist only if $A$ is a $\mathrm{PID}$. 
In a $\mathrm{PID}$, strong Gr\"obner basis coincides with the short reduced Gr\"obner basis. 
We give below an example to illustrate the difference between strong Gr\"obner basis  and short reduced Gr\"obner basis.
\begin{example}
Consider the polynomial ring, $\Bbbk[a_1,a_2][x]$. Let $G= \{ a_1^{2}x, a_2^{2}x\}$ be the set of generators of an ideal in  $\Bbbk[a_1,a_2][x]$.
$G$ is the short reduced Gr\"obner basis for the ideal since $\{a_1^{2}, a_2^{2}\}$ is the reduced Gr\"obner basis of the ideal it generates in $\Bbbk[a_1,a_2]$. 
Consider $f = (a_1^{3} + a_2^{3} )x$. We have, $\mathrm{lt}(G) = \{a_1^{2}x, a_2^{2}x\}$ and $\mathrm{lt}(f) = (a_1^{3} + a_2^{3} )x$. 
There exists no $g \in G$ such that $\mathrm{lt}(g) \mid \mathrm{lt}(f)$. But, $\mathrm{lt}(f) \in \langle\mathrm{lt}(G)\rangle$. Therefore, $G$  is not a strong Gr\"obner basis. 
\end{example}
\section{Border Basis over Residue Class Rings over Rings}
\label{borderbases}
\noindent Border bases have proven to be a numerically more stable tool to describe zero-dimensional ideals in $\Bbbk[x_1,\ldots,x_n]$ than Gr\"obner bases \citep{Kreuzer:2005:borderbases}. 
 In this section, based on the characterization given in Proposition ~\ref{Proposition for characterization},
we extend border bases to polynomial rings over the ring $A$. 

We can define the order ideal and the border of an order ideal in the same way as we have in  $\Bbbk[x_1,\ldots,x_n]$. 
\begin{definition}
A finite set $\mathcal{O} \subseteq {\mathbb{Z}}_{\ge 0}^n$ is called an order ideal if $x^\alpha \in \mathcal{O}$ and $x^\beta \mid x^\alpha$ where $x^\beta \in {\mathbb{Z}}_{\ge 0}^n$
implies $x^\beta \in \mathcal{O}$. 
\end{definition}
\begin{definition}
Let $\mathcal{O} \subseteq {\mathbb{Z}}_{\ge 0}^n$ be an order ideal, then border of  $\mathcal{O}$ is defined as 
\begin{displaymath}
\partial\mathcal{O} = (x_1\mathcal{O}\cup x_2\mathcal{O} \cup \cdots \cup x_n\mathcal{O})\setminus \mathcal{O}
\end{displaymath}
\end{definition}
We now define the $\mathcal{O}$- border prebasis. The only difference we have here is that the coefficients come from the ring $A$.
\begin{definition}
Let $\mathcal{O} \subseteq {\mathbb{Z}}_{\ge 0}^n$, $\mathcal{O} = \{x^{\alpha_1}, \ldots, x^{\alpha_s}\}$ be an order ideal and 
$\partial\mathcal{O} = \{x^{\beta_1}, \ldots, x^{\beta_t}\}$ be the border of $\mathcal{O}$. 
Then a finite set of polynomials $\mathcal{B} =\{b_1, \ldots, b_t\} \subseteq A[x_1,\ldots,x_n]$ is said to be $\mathcal{O}$- border prebasis if $\{b_1, \ldots, b_t\}$ are of the form,
\begin{displaymath}
b_i = x^{\beta_i} - \sum_{j=1}^s c_{ij} x^{\alpha_j}, c_{ij} \in A. 
\end{displaymath} 
Given an ideal $\mathfrak{a} \subseteq A[x_1,\ldots,x_n]$ if $\mathcal{B} \subseteq \mathfrak{a}$ then $\mathcal{B}$ is said to be an $\mathcal{O}$-border prebasis of $\mathfrak{a}$. 
\end{definition}
We give below the definition of border basis. Once we have an $\mathcal{O}$-border prebasis with coefficients from the ring the definition of  border basis directly follows.  
\begin{definition}
Let $\mathcal{O} \subseteq {\mathbb{Z}}_{\ge 0}^n$, $\mathcal{O} = \{x^{\alpha_1}, \ldots, x^{\alpha_s}\}$ be an order ideal and $\mathcal{B} = \{b_1, \ldots, b_t\} \subseteq A[x_1,\ldots,x_n]$ be
an $\mathcal{O}$-border prebasis. 
Let $\mathfrak{a}\subseteq A[x_1,\ldots,x_n]$ be an ideal such that $A[x_1,\ldots,x_n]/\mathfrak{a}$ is finitely generated and  has a free $A$-module representation w.r.t. $G$. 
Then $\mathcal{B}$ is said to be an $\mathcal{O}$-border basis if 
$ \mathcal{B}\subseteq \mathfrak{a}$ and $\mathcal{O} = \{ x^\alpha : \mathrm{lm}(g) \nmid x^\alpha, \forall g \in G, \text{ where } G \text{ is a monic short reduced Gr\"obner basis of }\mathfrak{a}\}$.  
\end{definition}
We give below certain results associated with border basis in $\Bbbk[x_1,\ldots,x_n]$ that are valid in $A[x_1,\ldots,x_n]$ as well. 
The proofs of these theorems are exactly in the same lines as in $\Bbbk[x_1,\ldots,x_n]$ and hence we skip them here. 
Note that in fields, border basis is defined only for zero-dimensional ideals. In rings, $\mathcal{O}$-border basis exists if and only if $A[x_1,\ldots,x_n]/\mathfrak{a}$ is 
finitely generated and free, i.e. if a short reduced Gr\"obner basis of $\mathfrak{a}$ is monic. 

We have the following theorem that illustrates how border bases generalize the notion of Gr\"obner bases.
\begin{theorem}
Let $\mathcal{O} = \{x^{\alpha_1}, \ldots, x^{\alpha_s}\}$ be an order ideal and $\mathcal{B} \subseteq A[x_1,\ldots,x_n]$ be an $\mathcal{O}$-border
basis of an ideal $\mathfrak{a} \subseteq A[x_1,\ldots,x_n]$. 
Then $\mathcal{B}$ generates $\mathfrak{a}$.
\end{theorem}
Uniqueness of the $\mathcal{O}$-border basis can be extended easily to rings.
\begin{theorem}
Let $\mathcal{O} = \{x^{\alpha_1}, \ldots, x^{\alpha_s}\}$ be an order ideal and let $\mathfrak{a} \subseteq A[x_1,\ldots,x_n]$ be an ideal such that 
$A[x_1,\ldots,x_n]/\mathfrak{a}$ is finitely generated and 
 has a free $A$-module representation w.r.t. $G$. 
Assume that 
$\mathcal{O} = \{ x^\alpha : \mathrm{lm}(g) \nmid x^\alpha, \forall g \in G, \text{ where } G \text{ is a }\newline \text{monic short reduced Gr\"obner basis of }\mathfrak{a}\}$.
Then there exists an unique $\mathcal{O}$-border basis, $\mathcal{B}$ of $A$.
\end{theorem}
The below theorem illustrates the uniqueness of the remainder when we reduce it with an $\mathcal{O}$-border basis. 
\begin{theorem}
Let $\mathcal{B} = \{b_1, \ldots, b_t\} \subseteq A[x_1,\ldots,x_n]$ be an $\mathcal{O}$-border basis of an
ideal $\mathfrak{a} \subseteq A[x_1,\ldots,x_n]$. Let $f \in A[x_1,\ldots,x_n]$, then remainder of $f$ when we reduce it with $\mathcal{B}$ is unique.  
\end{theorem}
\begin{example}
Let $\mathcal{O} = \{1,x\}$ be an order ideal in $\mathbb{Z}[x,y]$. The border of $\mathcal{O}$, $\partial\mathcal{O} = \{x^2, y, xy \}$. 
Then $\mathcal{B} = \{x^2 - 1, y -1, xy - x\}$ is an $\mathcal{O}$ - border prebasis. Consider the 
ideal $\mathfrak{a}$ generated by $\{x^2 - 1, y -1, xy - x\}$. Clearly, $\mathcal{B} \subseteq \mathfrak{a}$. 
Consider, $\mathbb{Z}[x,y]/\mathfrak{a}$. First, we have to determine if it is free and we use our characterization result for that. Then, we need to determine if the order ideal $\mathcal{O}$ is a 
$\mathbb{Z}$-module basis of the quotient ring.  Consider the generator set $\mathcal{B} = \{x^2 - 1, y -1, xy - x\}$. 
A Gr\"obner basis of this ideal is $\{x^2-1, y-1\}$. Since it is monic, $\mathbb{Z}[x,y]/\mathfrak{a}$
is free. The order ideal $\mathcal{O} = \{1,x\}$ forms a $\mathbb{Z}$-module basis of $\mathbb{Z}[x,y]/\mathfrak{a}$. 
This means $\mathcal{B} = \{x^2 - 1, y -1, xy - x\}$ is an $\mathcal{O}$-border basis of the ideal, $\mathfrak{a}$. 
\end{example}

 \section*{Summary}
 In this paper, we introduced the concept of short reduced Gr\"obner bases with which we characterized a finitely generated, free residue class polynomial ring over a ring, $A$ w.r.t a monomial order. 
The characterization identifies the residue class polynomial rings in $\mathbb{Z}[x_1, \ldots, x_n]$ for which all ideals in them are lattices (subgroups of $\mathbb{Z}^N$). 
The characterization gives rise to an algorithm, that uses short reduced Gr\"obner bases, to compute an $A$-module basis for a finitely generated, free residue class polynomial ring over $A$. 
Using this, we show that the concept of border bases can be extended to polynomial rings over rings.
 
\footnotesize{\section*{Acknowledgments}
The authors would like to thank the editor and the anonymous reviewers for their suggestions that resulted in improving the quality of the paper. We thank the anonymous reviewer for pointing out 
that the $A$-module surjective homomorphism in Section 3.2 is also an isomorphism.  }

\end{document}